\documentclass[sigconf]{acmart}
\usepackage[titletoc,title]{appendix}
\usepackage[export]{adjustbox}
\usepackage{tabularx}
\usepackage[most]{tcolorbox}
\usepackage{lipsum}  
\usepackage{graphicx}
\usepackage{pifont}
\usepackage{subcaption}
\usepackage{float} % for precise figure placement
\usepackage{ulem}
\usepackage{enumitem}
\usepackage{xcolor}
\usepackage{soul}
\definecolor{lightblue}{rgb}{0.635, 0.749, 0.878}
\usepackage{graphicx}
\usepackage{float}
\usepackage{tabularx}
\usepackage{pifont}
\usepackage{subcaption}
\usepackage{float} % for precise figure placement
\usepackage{colortbl} % to add colors to the table
\usepackage{adjustbox}
\usepackage{censor}

\newcommand{\neutralaccounts}{

    \begin{table*}[h]
    \centering
    \fontfamily{cmss}\selectfont 
    \begin{tabular}{l l l l}
    \toprule
        \textbf{Sports} & \textbf{Photography} & \textbf{Recipes and Cooking} & \textbf{Pets and Animals} \\
        \toprule
        @BleacherReport & @StormHour & @RecipesIdeas & @ShouldHaveCat \\
        @SportsCenter & @photography & @FoodNetwork & @catshouldnt \\
        @espn & @PopPhoto & @Fooddotcom & @contextdogs \\
        @NFL & @natgeophotos & @FoodPornPhotos & @contextraccoons \\
        @sports & @photoshop & @Allrecipes & @pets \\
        @olympics & @thephotohour & @cookingchannel & @petsmart \\
        @mlb & @icphotog & @my\_recipes & @adoptapetcom \\
        @nba & @worldphotoorg & @recipeland & @animalplanet \\
    \bottomrule
    \end{tabular}
    \caption{We selected four categories of socially and politically neutral topics: sports, photography, recipes, and pets. The poster accounts followed between 10 to 15 random accounts from this list.}
    \label{tab:neutral-accounts}
    \end{table*}
}

\newcommand{\reporteraccounts}{
\begin{table*}[t]
\centering
\fontfamily{cmss}\selectfont 
\begin{tabular}{>{\centering\arraybackslash}m{0.15\linewidth}
                >{\centering\arraybackslash}m{0.1\linewidth}
                >{\centering\arraybackslash}m{0.18\linewidth}
                >{\centering\arraybackslash}m{0.15\linewidth}
                >{\arraybackslash}m{0.25\linewidth}}
\toprule
Reporter name & AI face & X username & Job & Description of photo \\
\midrule
Jessica W. & \includegraphics[width=\linewidth, height=1.5cm]{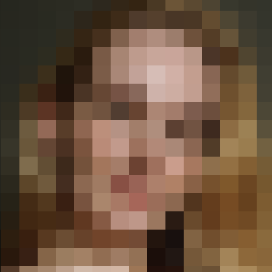} & @Jessica\censor{12345} & Student & A nude photo showing a woman smiling\\
\midrule
Ashley S. & \includegraphics[width=\linewidth, height=1.5cm]{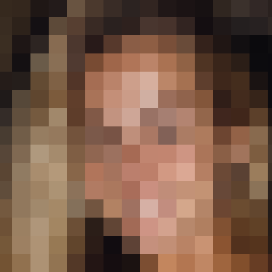} & @Ashley\censor{12345} & Student & This photo depicts a nude blonde woman seated \\
\midrule
Brittney J. & \includegraphics[width=\linewidth, height=1.5cm]{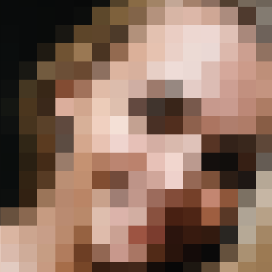} & @Brit\censor{123459} & Administrative assistant & Selfie of a blonde, smiling at the camera, naked\\
\midrule
Amanda K. & \includegraphics[width=\linewidth, height=1.5cm]{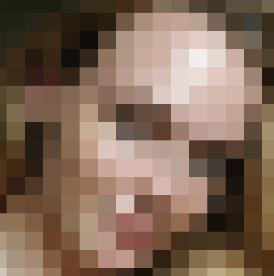} & @Amanda\censor{1234995} & Associate & This photo shows a woman with freckles, nude\\
\midrule
Emily B. & \includegraphics[width=\linewidth, height=1.5cm]{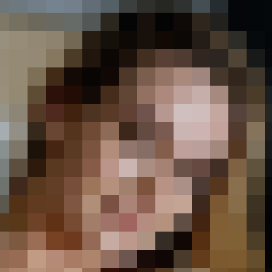} & @Emily\censor{12345} & Student & This is a photo of a blonde woman on a bed \\
\bottomrule
\end{tabular}
\caption{Summary of reporter accounts used to report content on X and information used for DMCA reporting. We selected common names for the reporter accounts. X usernames are structured with the first name and a string of numbers for all accounts. Job and description of photo are required fields for DMCA reporting. }
\label{reportersummary}
\end{table*}
}

\newcommand{\pixels}{
\begin{figure*}[h]
        \centering
        \includegraphics[trim = 0 110 0 0 , clip, width=\linewidth]{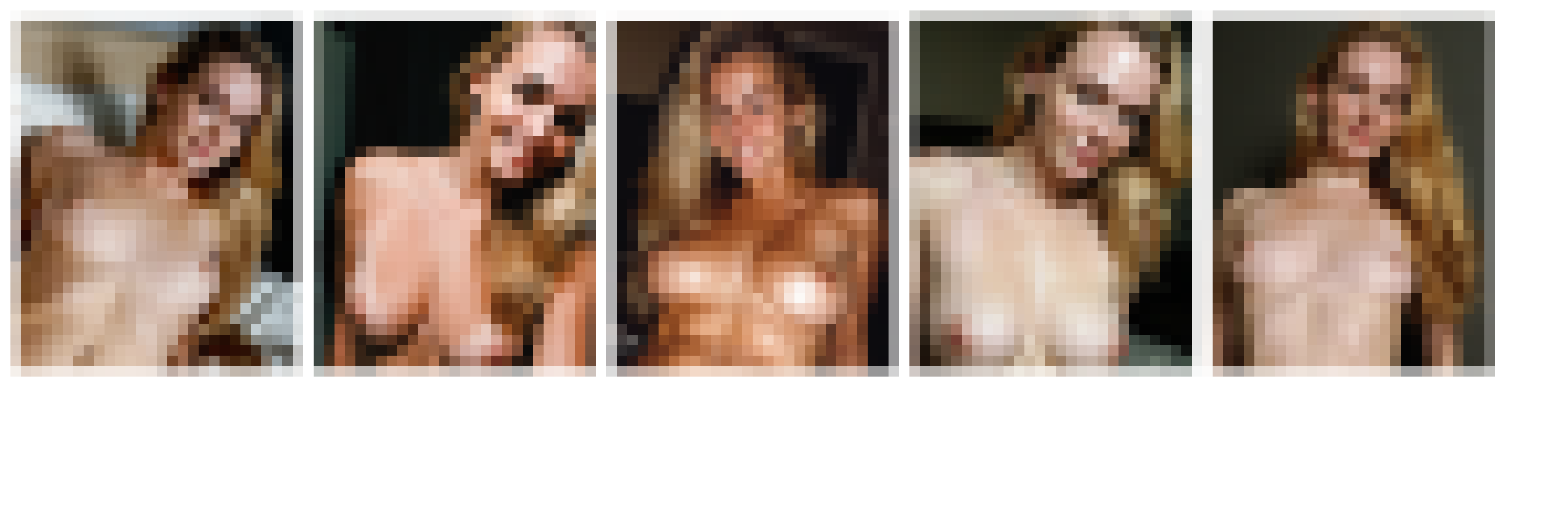}
        \caption{Pixelated versions of the five deepfakes we created, posted, and reported. We confirmed (via Google reverse image search, Yandex reverse image search, and PimEyes) these faces do not match any of individuals online. The deepfakes are meant to represent white women between the ages of 25 and 35, with a lighter hair color, depicted ``topless''. Images used in the study were not pixelated.}
        \label{fig:ncim-short-report}
    \end{figure*}
}

\newcommand{\twitterreporttable}{
\begin{table}[H]
\centering
\fontfamily{cmss}\selectfont 
\begin{tabularx}{\columnwidth}{>{\centering\arraybackslash}X>{\centering\arraybackslash}X>{\centering\arraybackslash}X}

     \textbf{Required Information} & \textbf{Copyright infringement} & \textbf{Non-consensual nudity}\\
\toprule
     Full name & \ding{52} & \ding{52}\\
     Street address & \ding{52} & \ding{56}\\
     Email address & \ding{52} & \ding{52}\\
     Official ID & \ding{56} & \ding{52}\\
     Infringement desc & \ding{52} & \ding{52}\\
     Infringing URL & \ding{52} & \ding{52}\\
     Original URL & \ding{52} & \ding{56}\\
     Legal ack & \ding{52} & \ding{56}\\
     Signature & \ding{52} & \ding{52}\\
\bottomrule
\end{tabularx}
\caption{Comparison of information required for reporting copyright infringement and non-consensual nudity on X.}
\label{table:twitterreporttable}
\end{table}
}

\newcommand{\ncimshortreport}{
    \begin{figure*}[H]
        \centering
        \includegraphics[width=0.9\linewidth]{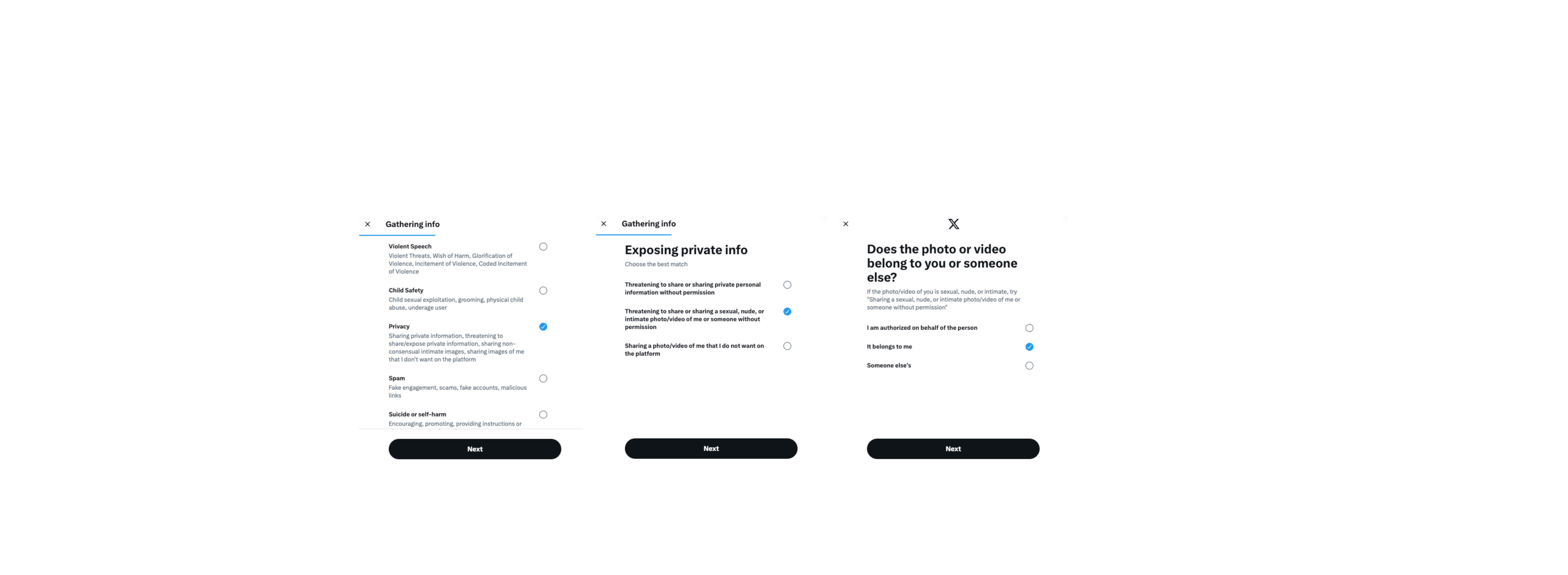}
        \caption{User interface for NCIM (private information) reporting option. These screens appear after clicking the \ldots icon of a post.}
        \label{fig:ncim-short-report}
    \end{figure*}
}

\setulcolor{lightblue} 
\setul{}{1.5pt}  

\AtBeginDocument{%
  \providecommand\BibTeX{{%
    \normalfont B\kern-0.5em{\scshape i\kern-0.25em b}\kern-0.8em\TeX}}}

\setcopyright{acmcopyright}
\copyrightyear{2024}
\acmYear{2024}
\acmDOI{XXXXXXX.XXXXXXX}

\acmConference[Conference acronym 'XX]{June 03--05, 2025}{TBD, TBD}

\acmPrice{15.00}
\acmISBN{978-1-4503-XXXX-X/18/06}

\begin{document}

\title{Reporting Non-Consensual Intimate Media: An Audit Study of Deepfakes}

\author{Li Qiwei}
\email{rrll@umich.edu}
\affiliation{%
  \institution{University of Michigan}
  \city{Ann Arbor}
  \country{USA}
}

\author{Shihui Zhang}
\email{zshihui@umich.edu}
\affiliation{%
  \institution{University of Michigan}
  \city{Ann Arbor}
  \country{USA}
}

\author{Andrew Timothy Kasper}
\email{atkasper@umich.edu}
\affiliation{%
  \institution{University of Michigan}
  \city{Ann Arbor}
  \country{USA}
}

\author{Joshua Ashkinaze}
\email{jashkinaze@umich.edu}
\affiliation{%
  \institution{University of Michigan}
  \city{Ann Arbor}
  \country{USA}
}

\author{Asia A. Eaton}
\email{aeaton@fiu.edu}
\affiliation{%
  \institution{Florida International University}
  \city{Miami}
  \country{USA}
}

\author{Sarita Schoenebeck}
\email{yardi@umich.edu}
\affiliation{%
  \institution{University of Michigan}
  \city{Ann Arbor}
  \country{USA}
}

\author{Eric Gilbert}
\email{eegg@umich.edu}
\affiliation{%
  \institution{University of Michigan}
  \city{Ann Arbor}
  \country{USA}
}

\renewcommand{\shortauthors}{Qiwei, et al.}

\begin{abstract}

Non-consensual intimate media (NCIM) inflicts significant harm. Currently, victim-survivors can use two mechanisms to report NCIM---as a non-consensual nudity violation or as copyright infringement. We conducted an audit study of takedown speed of NCIM reported to X (formerly Twitter) of both mechanisms. We uploaded 50 AI-generated nude images and reported half under X's ``non-consensual nudity'' reporting mechanism and half under its ``copyright infringement'' mechanism. The copyright condition resulted in successful image removal within 25 hours for all images (100\% removal rate), while non-consensual nudity reports resulted in no image removal for over three weeks (0\% removal rate). We stress the need for targeted legislation to regulate NCIM removal online. We also discuss ethical considerations for auditing NCIM on social platforms.

\end{abstract}

\begin{CCSXML}
<ccs2012>
   <concept>
       <concept_id>10003120.10003130.10003233</concept_id>
       <concept_desc>Human-centered computing~Collaborative and social computing systems and tools</concept_desc>
       <concept_significance>500</concept_significance>
       </concept>
   <concept>
       <concept_id>10003120.10003130.10011762</concept_id>
       <concept_desc>Human-centered computing~Empirical studies in collaborative and social computing</concept_desc>
       <concept_significance>300</concept_significance>
       </concept>
 </ccs2012>
\end{CCSXML}

\ccsdesc[500]{Human-centered computing~Collaborative and social computing systems and tools}
\ccsdesc[300]{Human-centered computing~Empirical studies in collaborative and social computing}

\keywords{Online sexual abuse, Content moderation, Twitter, X, Audit, Image-based sexual abuse, IBSA, Non-consensual intimate imagery, NCII}

\received{18 September 2024}

\maketitle

\section{Introduction}

\textit{\textcolor{orange}{Content warning: This text contains imagery and discussions of online sexual abuse.}}

As personal and intimate lives become increasingly mediated by social technologies, the issue of non-consensual intimate media (NCIM) demands urgent attention. NCIM includes so-called ``revenge pornography'' and sexualized or nude ``deepfakes'', and is also known as image-based sexual abuse (IBSA) and non-consensual intimate imagery (NCII). It involves the creation, obtainment, and (threats of) distribution of sexually explicit content without the depicted individual's consent. Alarmingly, NCIM is widespread: in the United States, 1 in 8 adults have had their intimate content shared without consent or have been threatened with such actions.\footnote{\url{https://cybercivilrights.org/research/}} Victim-survivors---90\% of whom are women---experience significant emotional distress and social withdrawal~\cite{eaton20172017,ccri2014revenge}. Tragically, some have taken their own lives or been killed as a direct result of NCIM~\cite{batool2024expanding,sambasivan2019they}.

A key concern for victim-survivors is the removal of abusive content from online platforms, which may circulate on both pornographic and mainstream social media sites~\cite{eaton_victim-survivors_nodate,mcglynn_its_2021,henry_image-based_2019,ccri2014revenge}. Reporting and removing NCIM is a challenging process. Although specific cases vary, there are some common experiences. First, the victim-survivor must somehow become aware that the abuse has occurred. If it is the first incident, she may discover it through a casual online search, from friends, or even from strangers on the internet. If the abuse is ongoing, she may use a paid service to scan for her content, hire legal representation, or receive alerts from monitoring websites. These activities are invariably traumatic~\cite{chen2022trauma, scott2023trauma}. Second, if she reports the content herself, she will need to visit the website hosting the content and search for removal options. This might involve filing a report under the Digital Millennium Copyright Act (DMCA), using the platform’s internal reporting system, emailing the site’s webmaster, or in some cases, finding that no contact method is available. Third, if she finds a way to report, she must choose between the platform’s internal reporting method or the DMCA, provided that she owns the copyright to the content~\cite{de_angeli_reporting_2021,ccri_online_removal}. A victim-survivor may hold the copyright if she captured the content herself (e.g. took the photo; filmed the recording). If using the DMCA, she must provide her full name, address, and a description of the infringement. If she does not own the copyright, she must resort to contacting the web host by any available means, whether through email or a contact form, if either is made available. Often, these messages are desperate pleas for removal. Finally, she waits. If fortunate, she may receive a response in days, weeks, or months. Unfortunately, NCIM content often remains online for years without being addressed~\cite{citron2022fight}.

Gender-based violence, including NCIM, has been a persistent societal harm. Prior literature on sexual assault reporting has detailed the systemic duress and lack of clarity faced by victim-survivors during the reporting process~\cite{eaton_victim-survivors_nodate,d2015fighting,de_angeli_reporting_2021,bedera2021settling}. In the realm of online sexual abuse, generative AI and its usage for sexualized deepfakes exacerbate the problem, now granting the ability for anyone to turn a photo of a face into abusive content~\cite{ohman_introducing_2020, maddocks_deepfake_2020, jacobsen_deepfakes_2024, de_ruiter_distinct_2021}. This paper investigates the efficacy of two mechanisms for removing NCIM: the DMCA, treating the image as a copyright infringement, and platform non-consensual nudity privacy policy. Intellectual property laws, including copyright and its internet counterpart, the DMCA, are intended to protect creative works and are enforced through federal legislation. In contrast, privacy is recognized as a human right, yet no laws mandate online platforms to remove NCIM~\cite{solove2008understanding}. Understanding how these mechanisms operate in practice is crucial for evaluating the principles they reflect. This study takes place on X (formerly Twitter), which is a private company owned by Elon Musk. %Since then, X has seen an expansion of free speech, and despite less engagement with content moderation, we find that X addresses with DMCA removal requests. 

We audit how different reporting methods---DMCA claims versus X's internal non-consensual nudity policies---affect the content removal process. We used generative AI to create realistic nude ``deepfake'' images of AI-generated personas, which were then posted on X. We reported these images and systematically collected data on the moderation processes, timelines, and outcomes for 21 days. We designed the study to minimize risks to users and content moderators by limiting the reach of the posts and by selecting a platform where there would be no volunteer moderators and little impact on paid moderators, if any.

Our results reveal a stark contrast between the two reporting mechanisms. All 25 images reported under the DMCA were removed within 25 hours, with a 100\% success rate. In contrast, all 25 images reported under X's non-consensual nudity policy received no response in over three weeks, resulting in a 0\% removal rate. Additionally, the five accounts we reported under the non-consensual nudity condition faced no consequences or notifications from X. Meanwhile, all five accounts reported under the DMCA condition received temporary suspensions for copyright violation. This discrepancy underscores the significant impact of the underlying legal frameworks supporting the DMCA. The effectiveness of the DMCA over internal policies and its subsequent punitive actions for infringing users strongly suggests the need for NCIM-specific federal legislation. NCIM requires similar, dedicated laws that mandate online platforms to uphold individuals' rights. To summarize, this study makes the following contributions:

\begin{itemize}
    \item \textit{\textbf{X removes copyrighted content and does not remove non-consensual nudity content: }} Content reported under the DMCA is removed within 25 hours, while content reported under the non-consensual nudity policy remained unaddressed over the full 3-week duration of this study.
    \item \textit{\textbf{Policy recommendations for NCIM regulation: }} Protecting intimate privacy requires a shift from dependence on platform goodwill to enforceable legal regulations. 
\end{itemize}

\section{Related Research}\label{rr}

Content moderation for NCIM intersects with both policy and technical domains. In this section, we provide an overview of content moderation goals and practices, examine current strategies for addressing NCIM online, explore the application and limitations of the DMCA, and discuss the significance of audit studies in this research area.

\subsection{Addressing NCIM online}

Non-consensual intimate media (NCIM) is a form of online sexual abuse, encompassing technology-facilitated abuse (GB-TFA), image-based sexual abuse (IBSA), and non-consensual intimate imagery (NCII). Despite terminological differences, these concepts converge on a common issue: the creation, distribution, or threats to distribute sexually explicit content that violates someone's privacy, facilitated by technology~\cite{citron2014criminalizing,citron_criminalizing_2016,franks_drafting_2014,qiwei2024sociotechnical}. Consent---clear, informed, and voluntary agreement to engage in some action---can be violated in a variety of ways in NCIM: during content creation, obtainment, or dissemination~\cite{qiwei2024sociotechnical}. Some forms of NCIM include: 
\begin{enumerate}
    \item \textit{Creation:} sexual assault recordings, "up-skirts", non-consensual sexualized deepfakes, fabricated media
    \item \textit{Obtainment:} sharing initially consensually obtained or filmed content, sharing private custom commercial content
    \item \textit{Dissemination:} showing intimate to others without consent, sending content in a group chat, posting content on social media, impersonation
\end{enumerate}

NCIM is disseminated through social computing infrastructure: public online forums, search engines, content recommendation algorithms, and end-to-end encrypted private channels~\cite{citron2018sexual,qiwei2024sociotechnical,scheffler2023sok}. While hashing-based systems such as StopNCII\footnote{https://stopncii.org/} have been adopted by social media platforms to proactively detect and block NCIM, this solution is only partial because perceptual hashing is not resistant to edits to content, and edited versions may be reuploaded easily and bypass filters~\cite{struppek2022learning}. 

Recent developments in generative AI technologies exacerbate the problem. With ``deepfake'' technologies, any photo containing a face can be manipulated into non-consensual sexualized content, showing the depicted individual nude or performing sexual acts without their knowledge or consent~\cite{rini_deepfakes_2022,timmerman_studying_2023,ohman_introducing_2020}. This potential for widespread dissemination significantly increases the harm inflicted on victim-survivors, making it easier for such content to reach a broader audience and persist online indefinitely.

Prior research has established the prevalence of NCIM, as well as highlighting significant psychological and social harms on victim-survivors~\cite{ccri2014revenge, sambasivan2019they, eaton_victim-survivors_nodate}. Victim-survivors experience social isolation, stigmatization, and damage to personal relationships~\cite{ccri2014revenge}. In many cases, this extends to career and reputational harm, where NCIM has been used as a tool for coercion and blackmail~\cite{citron2014criminalizing,citron2018sexual, eaton20172017, ccri2014revenge,bates_revenge_2017}. Victim-survivors also endure the violation of having their personal information---such as full names, family members' information, addresses, and social media accounts---doxxed online~\cite{mcglynn_its_2021,take2024expect}. %Research, primarily in psychology, has explored perpetration patterns, linking the likelihood of sharing NCIM to other personal traits or factors~\cite{karasavva_real_2021, henry2024image}.

Although 48 U.S. states and Washington D.C. have enacted legislation addressing NCIM, these laws have significant limitations that hinder justice for victim-survivors. First, legislation is jurisdiction-specific, complicating the prosecution of online crimes when the perpetrator resides in another state or country. Second, legislation primarily focuses on punitive actions for the perpetrator (when able to be identified), and offers little recourse for removing harmful content from online platforms---a critical need for victim-survivors~\cite{mcglynn_its_2021}. In other words, while legal action under current state laws have made progress in holding perpetrators accountable, they have failed to support victim-survivors with legislation that empowers them to remove NCIM depicting them from the internet. Numerous websites exist solely to distribute NCIM and harass victims, further exacerbating the challenges of content removal and protection~\cite{franks_drafting_2014, citron2020internet}. This rapid duplication and dissemination of content across multiple platforms makes it nearly impossible to fully eliminate especially.

The understanding of NCIM within sociotechnical research is still in its early stages. Existing studies contribute insights into how people manage privacy in sexual sharing, the use of privacy technologies by sex workers, and implications of NCIM reporting UIs~\cite{qin2024did, hamilton_nudes_2022, geeng_usable_nodate,qiwei2024sociotechnical,de_angeli_reporting_2023}. However, critical questions still remain unanswered: What are the outcomes of current content moderation processes for NCIM? Do platforms overlook or ignore the problem of NCIM? How can we design systems and policies that effectively respond to the problem of NCIM?

Legal scholars have expressed support for using the Digital Millennium Copyright Act (DMCA) to protect against NCIM distribution, largely because it successfully addresses many gaps in the current reporting mechanisms~\cite{farries2019feminist,d2015fighting,franks_revenge_nodate}. Copyright law is designed to ``stimulate the creation and dissemination of creative and artistic works valued by society'' by giving copyright holders the right to prevent the unauthorized distribution of their creations~\cite{fromer2015should}. It primarily protects creations for financial or commercial purposes, providing a mechanism to remove duplicated content that affects the copyright holder's income. The DMCA mandates that online platforms ``promptly'' process and remove copyrighted material upon receiving valid takedown notices. In 2021 alone, over 150,000 DMCA takedown notices were filed on X, demonstrating the widespread use of this legal mechanism online~\cite{twitter_transparency_2022}.

Photos are considered copyrighted by the photographer. This means that some victim-survivors hold the copyright to their photos or videos. Unfortunately, there are considerable drawbacks: the DMCA does not cover photos taken by others, requires extensive submitter information, and the cost of using paid services for DMCA claims can be prohibitive for many victim-survivors~\cite{seng2021copyrighting,o2020using,citron2020internet}. Online platforms benefit from Section 230 protections, which shield them from legal responsibility for user-generated content, including NCIM~\cite{gilden_sex_nodate,us_copyright_office_digital_replicas_2024}. This leaves victim-survivors vulnerable, as platforms have no legal incentives to remove NCIM in the same way they must for copyrighted material. Finally, the usage of the DMCA outside of its intended commercial usage remains controversial among copyright scholars, with concerns that copyright laws leveraged to protect sexual privacy would ``distort the intellectual property system''~\cite{fromer2015should,gilden_sex_nodate}.

\begin{figure*}[t]
    \centering
    \includegraphics[trim = 10 220 700 80 , clip, width=0.9\linewidth]{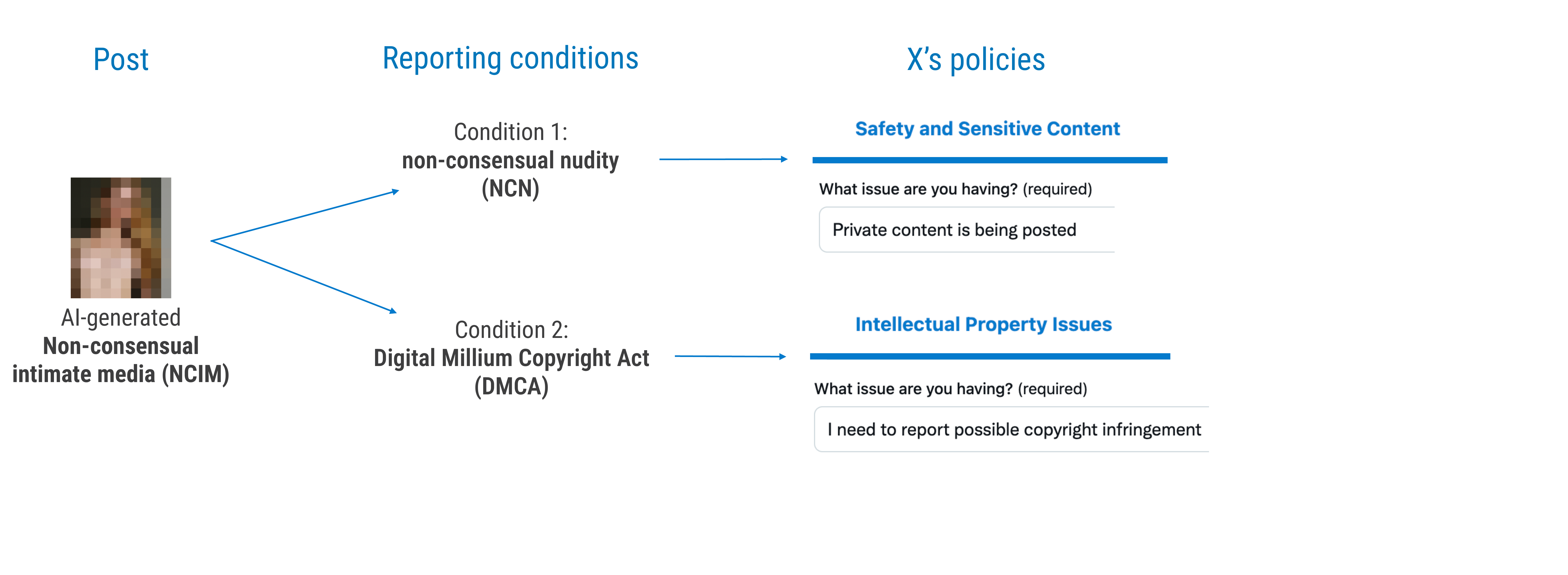}
    \caption{We audit two NCIM reporting mechanisms on X. In the first condition, we report under X's \textit{non-consensual nudity} policy, a \textit{privacy} report belonging to ``safety and sensitive content''. In the second condition, we use the DMCA, a \textit{copyright} report under ``intellectual property''.}
    \label{fig:reporting-flow}
\end{figure*}

\subsection{Content moderation and X}
Content moderation involves identifying and addressing problematic user-generated content online. Most online platforms employ a combination of automated systems and human moderators working together to enforce platform policies~\cite{jhaver_online_2018,thach2024visible}. Platforms define what is considered ``toxic'' or ``harmful,'' typically guided by legal policies and social norms. A broad range of topics are commonly studied and moderated, and while none have been fully ``solved''---as content moderation is a constantly evolving field---significant progress has been made in understanding, addressing, and mitigating these issues. This includes, but is not limited to:

\begin{enumerate}
    \item \textit{Spam:} Some of the earliest content moderation works are to limit advertisements, solicitations, and other broadcasts of unwanted content. Classification models are commonly used to distinguish useful content from spam~\cite{jindal2007review, rathod2015content,ndumiyana2013spam}.
    \item \textit{Misinformation:} Political misinformation rose as an important research topic during the prior decade~\cite{wilson2020cross, pennycook2019fighting}. Some popular methods include predicting truthful versus misleading news articles, suppressing viral content in networks, and a variety of crowd-sourced methods to identify and remove false information~\cite{stewart2021detecting, denaux2020towards, pennycook2019fighting}.
    \item \textit{Hate speech:} Social media platforms sanction users for remarks that are bullying, harassing, or otherwise harmful. Research has explored different types of penalties, community management strategies, and the usage of text models to predict hateful speech~\cite{schopke2023we, chandrasekharan2017you, warner2012detecting}. 
\item \textit{Child Sexual Abuse Material (CSAM):} Possession and distribution of CSAM is illegal in the United States~\cite{justice2023csam}. Due to its severity, CSAM is typically addressed with hashing-based databases such as PhotoDNA~\cite{farid2009image}. Content identified as CSAM is hashed, and stored in a database. When a user attempts to upload content, it is checked against the database before the post is made public to prevent the same content from appearing online~\cite{farid2009image,lee2020detecting, thiel2023child}.
\end{enumerate}

X, formerly known as Twitter, is a popular site for research and discussions on content moderation because it is widely used, involved in wide-ranging conversations, and relatively accessible for research. Prior work on X and its content moderation practices have examined a range of topics, including bot networks, the spread of political information, and interventions aimed at improving platform health~\cite{gillespie2018custodians,alizadeh2022content,im2020still}.

Despite extensive research on content moderation, there are limited empirical and quantitative studies to address NCIM moderation~\cite{mcdonald2021s,qin2024did,tseng2020tools,geeng_usable_nodate,batool2024expanding}. NCIM, which involves adults (18+), is managed differently from CSAM. Although both are illegal sexual content, CSAM has clearly defined boundaries: if it is CSAM, it should \textit{never} appear. However, because NCIM involves adults, it typically coexists with consensual adult content, making its provenance difficult to trace and its prevention challenging. Since Elon Musk took ownership of X in October 2022, the platform's priorities have shifted towards free speech, leading to the emergence and tolerance of a broader range of content~\cite{stanger2024first,arun2024x}. This shift, however, has also resulted in the proliferation of potentially harmful content. Against this backdrop, investigating X's content moderation practices becomes increasingly pertinent. In January 2024, X announced that users may ``share consensually produced and distributed adult nudity or sexual behavior'' on its platforms~\cite{x_adult_content}. However, identifying consensual content presents unique challenges: How can a platform determine if content is consensually created and distributed? If the content is reported to be non-consensual, how will this report be processed? Will the platform take steps to distinguish between content that is consensually recorded but non-consensual distributed? A large body of research and policy will be needed to determine how to effectively remove non-consensual content.

\subsection{Audit studies}

An audit is a method to systematically probe and evaluate a process, typically to identify discrimination~\cite{bertrand_are_nodate, sandvig2014auditing}. Audits are especially valuable when it is impractical or impossible to understand how a system operates from the outside looking in. This approach is valuable when the process being studied lacks transparency~\cite{sandvig2014auditing}. The social sciences have made significant strides in developing audit methodologies, including in-person audits of healthcare practices and resume audits to study bias against different demographic groups~\cite{paton2015clinical, bertrand_are_nodate}.

As AI systems began making decisions that impact people's lives, algorithmic auditing gained importance and relevance. Many algorithms function as ``black boxes'', making an audit a logical approach to uncovering their inner workings~\cite{sandvig2014auditing}. By systematically examining inputs and outputs, researchers can gain insights into how these algorithms operate without directly accessing code. Similar to traditional audit studies, the primary focus of algorithmic audits has been on detecting discrimination based on gender, race, and class~\cite{sandvig2014auditing}.  Auditing is both a method and a statement of values---deciding to conduct an audit asserts a belief that a system or entity should be subjected to scrutiny and high standards. Over the past decade, many critical decisions have become mediated by AI algorithms, making algorithmic auditing an important tool for sociotechnical researchers~\cite{sweeney2013discrimination,sandvig2014auditing}.

One of the earliest and most influential online audit studies found that names typically associated with Black Americans were 25\% more likely to trigger ads suggesting an arrest record compared to names associated with White Americans~\cite{sweeney2013discrimination}. This study raised critical early questions about racial discrimination in online spaces. Since then, other research has explored fairness in areas such as housing price discrimination ~\cite{asplund2020auditing, rosen2021racial} and in the online delivery of job ads~\cite{imana2021auditing}. More recently, Lam et. al expanded the audit framework by introducing the concept of the sociotechnical audit, which combines algorithmic auditing with user auditing. This approach not only examines the algorithms but also considers how system outputs affect users, offering a more comprehensive view of fairness and discrimination in technology-driven systems~\cite{lam_sociotechnical_2023}. While audits of content moderation are less well-established, the approach allows us to inspect moderation practices that are typically obfuscated.

\subsection{Research questions}

This study seeks to understand how reporting mechanisms influence the efficacy of NCIM removal. It asks the following research question:

\begin{tcolorbox}[colframe=lightblue, colback=white, boxrule=0.3mm]

    %\centering\textbf{Research Question}\\
    \textbf{How does the reporting mechanism---copyright vs. non-consensual nudity---impact NCIM removal?}

\vspace{0.8em}

\begin{description}
    \item[\textit{RQ 1: Image Removal from X:}] A boolean indicating whether the content was removed within a three-week (21-day) period after the report was made.
    \item[\textit{RQ 2: Hours to removal:}] If removed, the number of hours from the time of reporting until content is removed.
    \item[\textit{RQ 3: Other outcomes:}] The number of views AI-generated images collect on X. Punitive actions for accounts that posted the images. 
\end{description}
\end{tcolorbox}

\begin{figure*}[t]
  \includegraphics[width=\textwidth, trim={0 0 0 0}, clip]{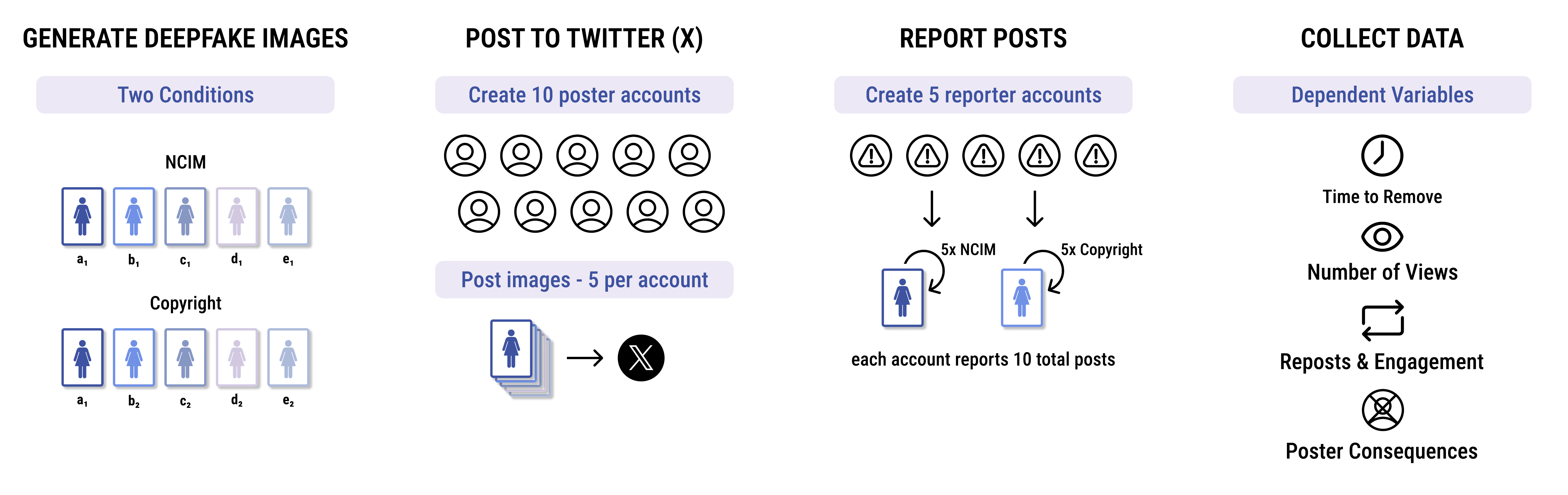}
  \caption{\textit{Method:} We created five unique AI-personas, and made a total of 50 posts on X via 10 new accounts called \textit{poster accounts}. We then reported 25 of the images as non-consensual nudity and 25 as DMCA violations. Finally, we collected data on the time to remove and other reporting outcomes.}
  \label{fig:hero}
\end{figure*}

We monitor for content removal up to three weeks (21 days) after the last report is made. We choose three weeks as a very liberal marker, considering that most of the harms are done in the first days after new content is posted. Online content peaks within 30 hours, while popular hashtags lose attention after 17 hours~\cite{salvania2015information, lorenz2019accelerating}. Both the TAKE IT DOWN Act and Meta’s Oversight Board mandate non-consensual media be removed within 48 hours~\cite{oversightboard_deepfake_2024,take_it_down_2024}. Given how content spreads online, three weeks is a sufficiently long window to capture meaningful removals.

\section{Method}\label{study-design}

We describe the design decisions, reporting conditions, materials, and measurements for the study. Figure ~\ref{fig:hero} depicts an overview.

\subsection{Design decisions}

To refine our study design, we conducted a rigorous set of pretests across distinct design areas.

\subsubsection*{Reporting methods}

X's guidelines note two ways to report non-consensual nudity. The first is to click the three horizontal dots on the post's UI, as shown in Figure \ref{fig:ncim-short-report} in the Appendix. The second way is through a separate reporting form, as shown in Figure \ref{fig:safety-form} in the Appendix. We evaluated both reporting methods and chose the latter form for several reasons. First, reporting forms exist for both DMCA (copyright) and NCIM (non-consensual nudity). In contrast, clicking three dots does not display an option for reporting copyright, so using the form to report reduces potential confounds for the reporting pipeline. Second, the reporting form includes a text box to compose a message for the content moderation team or system, which we integrated into our study design.

\subsubsection*{Watermarking photos}
Reporting copyright infringement requires a link to the original content. We explored including a visible watermark on the original content to see if it would lead to faster removal. A watermark that clearly indicates the copyrighted nature of a photo might lend more credibility to the DMCA report. In our pretest, we submitted copyright reports using both watermarked and unwatermarked versions of the same image. We found no difference in removal time. To eliminate any confounds with watermarked photos, we decided to link to an original image \textit{without} a visible watermark for all links to the original content.

\subsubsection*{Reporting duplicate photos}
We considered whether reporting one photo on X would also surface its duplicates. For example, if photos $P_1$ and $P_2$ are duplicates, would submitting a copyright report for $P_1$ also result in the removal of $P_2$? We tested this in two scenarios. First, when $P_1$ is posted by account $A$ and $P_2$ by account $B$, we examined whether reporting $P_1$ and account $A$ would affect $P_2$ and account $B$. Second, when both $P_1$ and $P_2$ are posted by account $A$, we investigated whether reporting $P_1$ would impact $P_2$. We found that neither scenario led to the removal of $P_2$. This means that even if the same image is posted multiple times by the same account, reporting one instance only results in the removal of that single image. %These results suggest that X does not use hashing-based algorithms to manage DMCA reports, as such algorithms would likely detect and remove duplicate content, at least at the account level.

\subsubsection*{Crafting the DMCA message}

Because we opted to use the form version for both DMCA and non-consensual nudity reports, we also constructed messages that appropriately described the DMCA and non-consensual nudity infringements. These two messages were designed to be as similar as possible, while still remaining suitable for each reporting condition. A key difference between the two messages is that the DMCA message serves as a legal statement, which requires specific language that does not apply to non-consensual nudity reports. Pretesting revealed that a brief DMCA message failed to result in content removal; we were asked to provide additional information. We extended the DMCA reporting message to meet the required standards. The messages used are provided in Appendix \ref{dmca-message}.

\subsection{Reporting conditions}\label{method-conditions}

The ``standard'' use of the DMCA is to remove unauthorized publications of content like licensed novels, movies, music, and other works to protect the rights-holder's financial and commercial interests. The DMCA can also be leveraged to protect privacy by removing NCIM. %Despite this use, leveraging the DMCA for NCIM is controversial, not always accessible, and can inadvertently expose reporters to additional harms.

The key difference between the DMCA and X's non-consensual nudity policy lies in their legal foundations, or lack thereof. The DMCA is a federal law that platforms must comply with, while the non-consensual nudity policy is X's own internal guideline, with no oversight from external governing bodies. Although both serve as avenues for content removal, they differ significantly in terms of enforcement and accountability. By comparing removal times, we can assess how rigorously X adheres to federal regulations versus its internal policies, where no external enforcement exists.

\reporteraccounts

\subsubsection*{NCIM} X's non-consensual nudity policy was last updated in December 2021~\cite{twitter_intimate_media_policy}. The non-consensual nudity policy prohibits posting or sharing intimate photos or videos produced or distributed without consent. including sexualized deepfakes. It mandates immediate and permanent suspension for original posters of non-consensual content. X's rules allow for exceptions for accounts that inadvertently repost or disseminate NCIM, highlighting the risk of unintentional sharing. We reported NCIM under the non-consensual nudity policy via the \textit{private information report form} which is accessible under ``safety and sensitive content'' on X's help page. Table \ref{table:twitterreporttable} in the Appendix outlines the information required to submit a report via the privacy form, and the full version of the form is provided in Appendix \ref{content-policy}. Table \ref{tab:terms} in the Appendix notes X's organization of different policies.

\subsubsection*{Copyright} X's copyright policy, accessible under the ``Intellectual Property'' section outlines the platform's process for responding to copyright complaints in accordance with the DMCA~\cite{x_copyright_policy}. The page offers comprehensive instructions on filing a copyright complaint, including the specific information required for submission. The information provided in a DMCA complaint, such as the reporter’s full name, email address, street address, and any other details included in the complaint, is shared with the user who posted the allegedly infringing content. The copyright policy directs users to an intellectual property form, where they can select the option ``I need to report possible copyright infringement'' to initiate the complaint process. A complete version is available in Appendix \ref{content-policy}.

\subsection{Creating deepfakes and accounts}\label{method-materials}

\subsubsection*{Sample size}
To determine the required sample size, we considered the effect size between NCIM and DMCA reports, using an $\alpha$ of 0.05 and a power of 0.8. Although pretesting suggested a very large effect size, we opted for a more conservative estimate of 0.4 for the power analysis. The initial analysis conducted in G*Power indicated a total sample size of 34, yielding an actual power of 0.801.

We use five unique photos, each representing an AI-generated persona. This selection of five images ensures that our study does not rely on a single image to represent all NCIM cases, which allows us to test the generalizability of our findings across different images and personas. The generated images are available in Appendix \ref{study-materials}. Each of the 5 unique photos is duplicated 10 times, providing five reports per photo under each of the two conditions, for a total of 50 images. %This approach enhances the study's rigor, ensuring that the observed effects are not confined to specific images or instances. Considering these factors, we determined a total sample size of 50, which further increases the study’s power. 

\subsubsection*{Deepfake personas} We used a generative AI model to create deepfake personas. We prompted the model to create five unique nude images depicting white women appearing to be in their mid-20s to mid-30s. Each image shows a woman nude from the waist up, including her face. These images were designed to represent realistic NCIM, which often involves younger women~\cite{ccri2014revenge}. Prior research suggests differential treatment for victim-survivors based on demographics such as race and gender~\cite{brubaker2017measuring, clark2012characteristics}. To minimize potential confounds from biased treatment, we chose to depict only younger, white women. Future research should explore differences in content removal for reporters of different skin tones and genders. Pixelated versions of the deepfake images are located in the Appendix \ref{study-materials}.

Generative models can inadvertently reproduce faces similar to those in their training data~\cite{tinsley2021face}. To prevent identity leakage and collision, we implemented a rigorous verification process and took precautions to ensure the ethical creation of these images. Each image was tested against a facial-recognition software platform and several reverse-image lookup services to verify it did not resemble any existing individual. Only images confirmed by all platforms to have no resemblance to individuals were selected for the study.

\subsubsection*{Poster accounts} We created 10 new X accounts, referred to as \textit{poster accounts}, to post the AI-generated images. We balanced making these accounts appear realistic while setting them up to be essentially identical to each other. We follow a rigorous protocol that controls for account names, age, photos, accounts followed, and engagement on X. Each account was created with a username in the format \textsc{@Firstname+[numbers]}, where \textsc{[numbers]} is a randomly generated string with length 3-10. The first names were randomly generated. All 10 accounts were created within a three-day period in 2024 to ensure they were similar in account age. During the creation of new accounts, X requires following accounts and selecting topics of interest. To meet these requirements while minimizing the potential for bias in content removal decisions, we created a bank of X accounts focused on politically and socially neutral topics: cooking, sports, photography, and pets~\cite{pennycook2021shifting}. See Appendix \ref{tab:neutral-accounts} for a full list of the neutral accounts. For each category, eight popular accounts were selected, making a total of 32 neutral accounts. Each poster account followed between 10 to 15 randomly chosen accounts from this bank. These neutral accounts constituted the entirety of each poster account's following list. To enhance realism, poster accounts engaged with the content they followed by liking or retweeting posts, but did not reply, comment, or post new tweets outside of the AI-generated images for the study. This approach minimized differences among the 10 accounts and limited their impact on existing social media communities. We allowed these accounts to naturally gain followers without further engagement. Considering the relatively short duration of these accounts being active and the significantly limited set of usage, there is minimal opportunity for the 10 poster accounts to differ from each other.

In summary, each poster account followed the following criteria during account setup: 
\begin{enumerate}
    \item\textit{Account name:} @Firstname + string of numbers
    \item\textit{Account age:} At the time of reporting, accounts were between 1-2 weeks in age
    \item\textit{Bio and profile photo:} No bio, no profile photo 
    \item\textit{Following:} Follow 10-15 random accounts from a designated bank of neutral, popular accounts
    \item\textit{Followers:} Poster accounts accumulated an average of 34.2 followers over three weeks
\end{enumerate}

\twitterreporttable

\subsubsection*{Reporter accounts} We created five X accounts to report the images, referred to as \textit{reporter accounts}. Each reporter account was associated with one of the five AI-generated personas and used a generic username in the format \textsc{@Firstname[number]}. Reporter accounts were not given profile photos or bios, and followed 2-3 suggested accounts during setup, as required by X. Though reporting behavior is not well understood, many reporters may decline to use their primary account for reporting due to the trauma associated with reporting~\cite{chen2022trauma,scott2023trauma}. Reporter accounts did not make posts or engage with other content on the platform. See Table \ref{reportersummary} for a summary of reporter accounts.

\subsection{Posting and reporting}\label{method-setup} 

\subsubsection*{Making posts} Soon after creation, each poster account made five posts over the course of two days, resulting in a total of 50 posts across 10 accounts. The posts were scheduled at random times between noon and midnight each day using X's post-scheduling feature. Each unique photo was posted once per account. To maintain uniformity and minimize potential confounds, all posts included the same set of hashtags without any additional captions. The hashtags---``\#porn'', ``\#hot'', and ``\#xxx''---reflect the most popular choices for adult sexual content on X. 

\subsubsection*{Reporting posts} We begin reporting posts under the non-consensual nudity condition 10 to 12 days after the posts are made. For logistical reasons, we waited an additional week before reporting under the DMCA condition. This makes the time between posting and reporting DMCA between 17 and 19 days. This difference is negligible for the analysis because time is calculated as the delta between reporting time and removal time, if removed. We chose to wait a sufficiently long time between posting and reporting, to allow X to potentially detect or remove this content without manual reporting. 

We reported posts from five accounts under the DMCA condition, and reported posts from the other five accounts under the non-consensual nudity condition. This setup allowed us to compare the potential punitive actions taken against accounts in each condition. We follow X's guidelines in reporting the content in each condition. While the reporting form offers an option to add more than one infringing URL with each report, we instead choose to create a separate report for each infringing URL. This choice is to increase the odds that these reports are processed by independent reviewers or machine instances. In total, we make 50 separate reports. Personal information, including job title, place of employment, and address is required to submit a DMCA report. To populate these fields, we randomly chose titles from common jobs such as ``associate'', ``student'', and ``administrative assistant''. The addresses used to report were also randomly generated from U.S.-based addresses. See Table \ref{table:twitterreporttable} for information required under each reporting condition.

Both ``non-consensual nudity'' and ``copyright infringement'' forms contain a text box for free user input. We use this space to communicate the removal request, crafting messages for the two conditions. Originally, we had wanted only to have the \ul{underlined} text differing between the two conditions. However, as discussed in Section \ref{method-setup}, DMCA messages require additional information. For this reason, we enrich the DMCA condition with added text following the statements below. For full reporting statements used in the messages, see Appendix \ref{dmca-message}. 

\begin{enumerate}
    \item \textit{DMCA/copyright condition:} \ul{I own the copyright to this photo.} I do not consent to it being posted. Please remove this immediately.
    \item \textit{NCIM/non-consensual nudity condition:} \ul{This is a photo of me.} I do not consent to it being posted. Please remove this immediately.
\end{enumerate}

To ensure consistency and avoid potential bias from policy changes or overwhelming X's moderation team, reports were submitted within a two-day window. This approach balanced the need for timely reporting while preventing an influx of labor for moderators.

\subsection{Measurements}\label{method-measurements}

The primary metric we capture is the time it takes for X to address reports and remove content, measured in hours from the moment the report is submitted. We set a generous cutoff time of three weeks (21 days); if the content is not removed within this period, it is recorded as not removed. This cutoff is based on legislation indicating that most harm occurs within the first 48 hours, with three weeks extending well beyond the window for effectively mitigating damage~\cite{take_it_down_2024, oversightboard_deepfake_2024}.

Additionally, we collect engagement data for each post and any punitive actions for reporter accounts. Victim-survivors are threatened by content that remains online. Content exposure, such as when content gains views, shares, and increases the chance that it would remain online or be seen by someone who knows her~\cite{qiwei2024sociotechnical,mcglynn_its_2021}. Warnings, suspensions, or messages from X would alert and inform posters of potential harms.

\begin{figure*}[t]
    \centering
    \includegraphics[trim = 200 110 400 10 , clip, width=\linewidth]{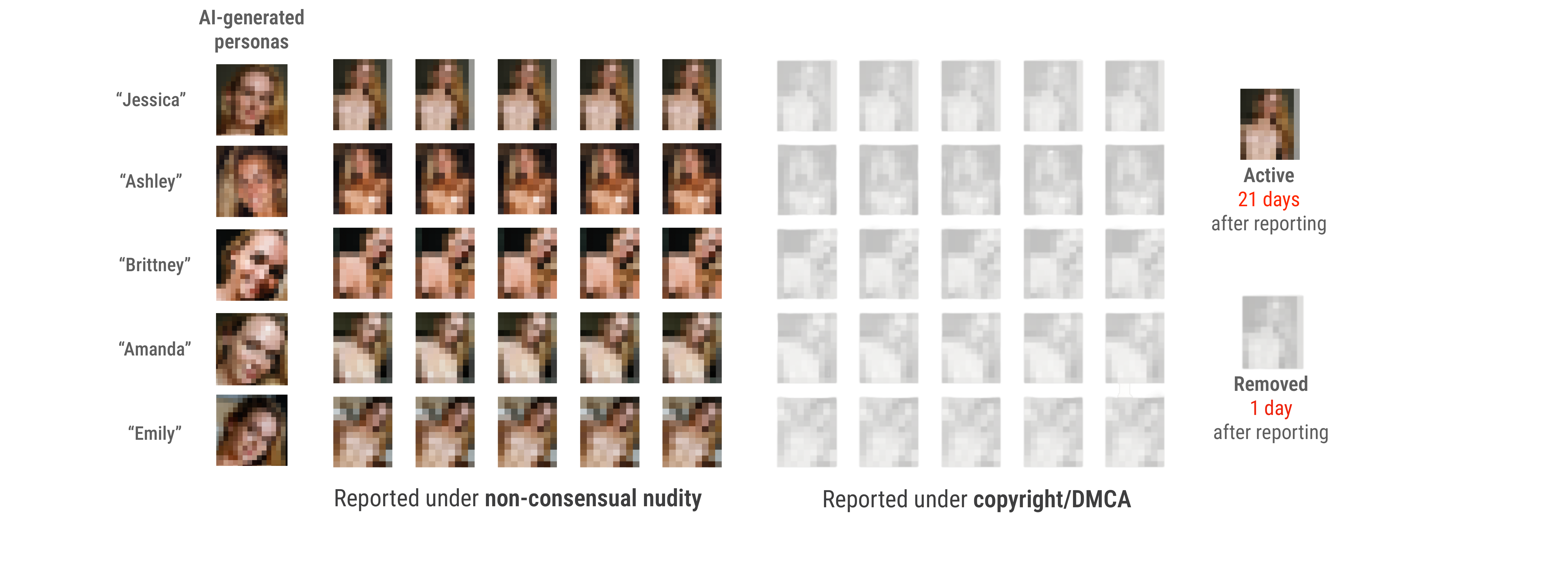}
    \caption{\textit{Findings: }25 NCIM images reported under DMCA/copyright were removed within approximately 25 hours after reporting. 25 NCIM images reported under X's non-consensual nudity policy were not removed for the 21-day duration of the study.}
    \label{fig:25-25}
\end{figure*}

To summarize, we collect the measures below: 
\begin{enumerate} 
    \item\textit{Removal within three weeks:} A boolean indicating whether the content was removed within the three-week period after the last report was made. 
    \item\textit{Hours to removal:} The number of hours from the time of reporting until content is removed. 
\end{enumerate}

Additionally, we note the following: 
\begin{enumerate}
    % these are QUALITATIVE CONFOUNDS - JUST PLOT DESCRIPVELY 
    \item\textit{Views and engagements:} The total number of views, likes, and comments each post receives. 
    \item\textit{Consequences for poster accounts:} Suspensions or sanctions imposed on the poster accounts after reports.
    % \item\textit{Reposts to other platforms:} Whether the photo was downloaded and reposted to other platforms, the names of these platforms, and the number of reposts.
\end{enumerate}

\section{Findings}\label{findings}

\subsection{RQ 1: Non-consensual nudity reports are not removed in 3 weeks}

We found a clear and substantial difference between the effectiveness of the two reporting mechanisms. All 25 reports submitted under the DMCA resulted in the successful removal of the NCIM content. In contrast, none of the 25 reports made under X’s non-consensual nudity policy led to the removal of the images within the three-week observation period. As a result, all images reported under the non-consensual nudity policy remained visible and active on the platform throughout the data collection period. Figure \ref{fig:25-25} provides a visual representation of these outcomes. 

\subsection{RQ 2: NCIM reported via DMCA reports are removed in under 25 hours}

\begin{figure*}[t]
    \centering
    \begin{subfigure}[b]{0.45\linewidth}
        \centering
        \includegraphics[width=\linewidth, trim={0 0 0 25}, clip]{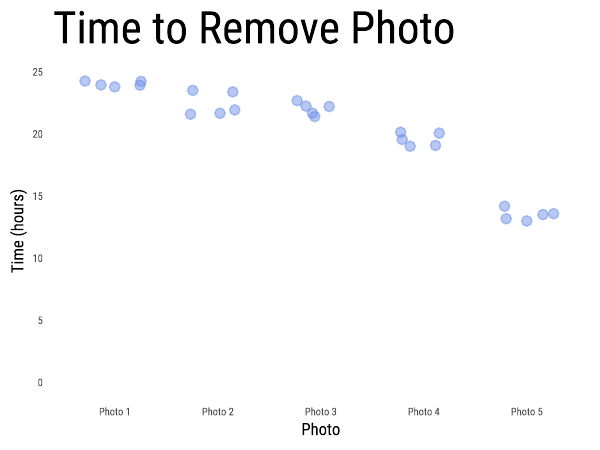}
    \end{subfigure}
    \hfill
    \begin{subfigure}[b]{0.45\linewidth}
        \centering
        \includegraphics[width=\linewidth, trim={0 0 0 25}, clip]{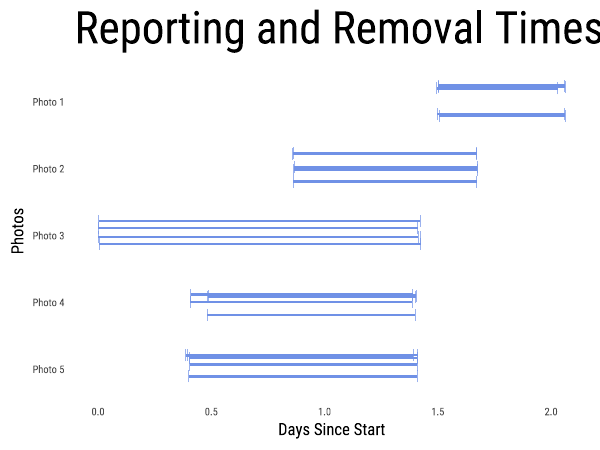}
    \end{subfigure} 
    \caption{Removal times for images reported under the DMCA condition, organized by the unique photo. The mean time for DMCA removals for all 25 images is 20.30 hours. The fastest case took approximately 13 hours.}
    \label{fig:combined-range-scatter}
\end{figure*}

For the DMCA reports, removal was prompt: all 25 reported images were taken down within 25 hours of reporting. As illustrated in Figure \ref{fig:combined-range-scatter}, the time to removal varied across the reports, with the fastest resolution occurring approximately 13 hours after the report was submitted. We submitted multiple reports for the same images around the same time to mirror the experience of a victim-survivor reporting their content in bulk. We observed that X's moderation team tends to address reports in ``batches''. All images related to a single AI-generated persona were removed within minutes of each other. This pattern suggests a possible operational approach where similar cases are handled collectively. The mean time for DMCA removals across all photos in that condition was 20.30 hours. The average number of hours to remove was 24.2 for photo 1, 22.61 for photo 2, 21.98 for photo 3, 19.43 for photo 4, and 13.3 for photo 5. 

\subsection{RQ 3: Negligible engagement on photos, no notifications to NCN poster accounts}
Posts that receive significantly more views than others may signal some differences in the poster account or the image itself. We collected engagement data for all 50 posts at the three-week mark, using the number of views as the primary measure. We found that all images received negligible views and no engagement in the form of likes, retweets, or comments. Across both DMCA and non-consensual nudity conditions, the average number of views over three weeks was 8.22, with a median of 7. Posts reported under the DMCA averaged 7.36 views, while NCN-reported posts averaged 9.08 views. There were no statistically significant differences in the number of views across the five unique photos. At the end of the three weeks, we also recorded the number of followers for each poster account. On average, poster accounts accumulated 34.2 followers each. The low number of views on the content posted is unsurprising given X's network structure. Content engagement is closely tied to the number of followers an account has. Since our newly created accounts had few followers, the posts naturally attracted minimal attention.

All five poster accounts for which we reported DMCA received temporary bans from X, and an email with information about the DMCA report. The DMCA report submitted by our reporter account is available in full in the email to the poster account. See both the suspended account and email in Appendix Figures \ref{fig:sus-account} and \ref{fig:sus-account-email}. All five NCN poster accounts did not receive any consequences, or notifications from X regarding these reports.

\section{Discussion}\label{discussion}

Our findings reveal a significant disparity in the effectiveness of content removal processes between reports made under the DMCA and those made under X's internal non-consensual nudity policy. Images reported for copyright infringement under the DMCA were removed within a day, while identical images reported under X's privacy policy remained on the platform for over three weeks. Two major changes are needed to address the problem of NCIM content: greater platform accountability, and the legal mechanisms to ensure that accountability.

\subsection{Platform accountability}

Establishing platform accountability requires that moderation decisions be more transparent and that the general public and experts can weigh in on those decisions. 

Benchmarks are useful for comparing against the state-of-the-art and are commonly in detecting hate speech and other unwanted behaviors online~\cite{poletto2021resources}. Could we imagine a benchmark for the speed of removing reported content? Similar to how first responders in medical emergencies operate within documented time frames, could platforms establish benchmarks for responding to different types of content reports? Furthermore, how can transparency be ensured without compromising the privacy and safety of individuals affected by NCIM? Often, content moderation introduces tensions between safety and privacy, which may be encountered in various moderation actions, values, and philosophies~\cite{jiang2023trade}.

\subsection{Policy implications}

Ultimately, protecting intimate privacy requires a shift from reliance on platform goodwill to enforceable legal standards. Relying on platforms to self-regulate user privacy has proven insufficient in the past. Tech platforms have a checkered history of privacy violations, as highlighted by repeated failures to adequately protect users' personal information, harvesting user data, harmful targeted advertising, and repeated data breaches~\cite{ke2023privacy, nissenbaum2011contextual, mayer2021now, ali2022all}. 

The stark contrast in removal outcomes highlights a critical gap in how NCIM is addressed through platform policies versus legally enforced mechanisms. While the DMCA benefits from robust federal backing, privacy policies related to NCIM on social media platforms lack the same legal muscle. Results from this study, combined with prior evidence of the DMCA's limitations for NCIM, strongly suggest the need for federally backed legislation that prioritizes privacy rights for non-consensual content as urgently as copyright. A dedicated NCIM law must clearly define victim-survivor rights and impose legal obligations on platforms to act swiftly in removing harmful content. Laws like the General Data Protection Regulations (GDPR) enacted by the European Union recognize that individuals ought to control their own data~\cite{albrecht2016gdpr,gdpr_what_is_gdpr}. While these laws cause disruptions to the existing data frameworks that online platforms have operated under, their calls to user consent and data privacy represent important steps forward.

\subsection{Ethics}

Our decision to conduct an audit of NCIM was not made lightly. While there were many study design decisions we made to minimize the likelihood and severity of harm (e.g. conducting the study on X rather than Reddit which has volunteer moderators who would have to review the content), there were two main ethical considerations in the study: the creation and public posting of deepfake nude content, and the use of DMCA requests outside of their intended use. 

\subsubsection*{Creating and posting deepfake sexual content}

As noted by Sandvig et. al, audit studies, by definition, cannot be conducted with informed consent, violating commonly-held ethical standards in experimental science~\cite{sandvig2014auditing}. This means that not everything should be an audit study---researchers must think carefully about the potential drawbacks and benefits. This study was classified as ``not regulated'' by our institution's review board because it technically does not involve human subjects; this outcome also highlights the limitations of relying on IRBs for ethical guidance. 

We took extensive precautions at each stage to minimize any additional harm. We confirmed to the best of our ability that the faces used in the models did not match real individuals. We did not attempt to boost our posts to gain additional exposure---which is often done with real NCIM content, but the benefits of doing that in our study did not outweigh the potential risks. 

There are multiple injustices involved in the creation of sexualized deepfake content. Nearly all (98\%) deepfake content online is pornographic, and models designed specifically for generating sexualized deepfakes are widely used and easily promoted by search engines and pornographic platforms~\cite{securityhero_state_of_deepfakes_2024}. The training data for these models is often collected without the consent of the individuals depicted, and it is impossible to determine how much of the original content was created consensually. These explicit deepfake models are easily accessible, allowing anyone to swap a person’s face into pornographic content. Additionally, due to potential data leakage, the faces created can sometimes match real individuals, causing inadvertent harm. We took these ongoing injustices into account when evaluating the ethical implications of using deepfake models in our study.

Our study may contribute to greater transparency in content moderation processes related to NCIM and may prompt social media companies to invest additional efforts to combat deepfake NCIM. In the long run, we believe the benefits of this study far outweigh the risks.

\subsubsection*{Use of DMCA}

We submitted DMCA requests for generative AI content in one condition. We weighed this choice carefully, recognizing that ethical audit testing sometimes requires breaching platform rules~\cite{aclu_sandvig_v_barr}. Our use of DMCA parallels Sandvig’s study, where researchers created artificial profiles to audit online hiring platforms for discrimination, a process that led to legal challenges of the Computer Fraud and Abuse Act (CFAA). Federal court rulings in favor of Sandvig's team affirmed the importance of such research. We also recognize that submitting DMCA requests as part of this study may impose an additional burden on the individuals processing these tickets. However, X received more than 150,000 DMCA reports in 2022 alone~\cite{twitter_transparency_2022}, and our 25 DMCA tickets represent a negligible increase---approximately 0.017\%. 

\subsection{Limitations and future work}

Though an audit should approximate real experiences closely, we are not able to capture the full range of NCIM harms in a safe design. One reason for this is due to X's networked design. In a networked setting, content posted by an account is pushed to its followers, cascading through the network. By using brand new accounts, we avoid the actual harms of disseminating NCIM but also do not accumulate engagement on the content we post. We are also not able to capture image downloads. Auditing the content moderation practices of a single company, rather than independent actions taken by different parties (such as various platforms). ties our findings to that particular context. However, all algorithm audits face the limitation of examining only one specific system or construct at a time~\cite{sandvig2014auditing}. Our audit was limited in size and scope, comprising only 50 total posts over a span of roughly four weeks from the beginning of posting to the end of data collection. Finally, our design and results are interpreted via one singular country's laws. Future work may explore policy implications for smaller and less regulated platforms, laws in a global context, and content moderation in the global majority~\cite{batool_expanding_2024}.

\section{Conclusion}

We found drastic differences between reporting NCIM under the DMCA and X's non-consensual nudity policy. Content reported under the DMCA is successfully removed while content reported under X's non-consensual nudity policy is not. This highlights the need for stronger and directed regulations and protocols to protect victim-survivors. Our study contributes to the broader understanding of platform accountability, content moderation processes, and how laws influence platform behavior. 

\section*{Acknowledgement}

We used OpenAI's ChatGPT 4o model to edit for text redundancies and suggest new word choices.

\bibliographystyle{ACM-Reference-Format}
\bibliography{2sample-base}

\newpage

\begin{appendices}

\section{Reporting text}\label{dmca-message}

\begin{enumerate}
    \item \textbf{Non-consensual nudity reporting text:} This is a photo of me. I do not consent to it being posted. Please remove this immediately.
    \item \textbf{DMCA reporting text:} I am the copyright owner of the content in question. I do not consent to it being posted. Please remove it immediately. I am writing to notify X of the unauthorized use of my copyrighted content. I hereby demand that X remove the tweet containing the infringing content. I make no representations or claims with respect to any third-party content that may incorporate or be included within the content. This notice does not purport to be a complete statement of the law or the facts and is without prejudice to my rights in the United States and abroad, which rights are expressly reserved. 512(f) Acknowledgement: I understand that under 17 U.S.C. § 512(f), I may be liable for any damages, including costs and attorneys' fees, if I knowingly materially misrepresent that reported material or activity is infringing. Good Faith Belief: I have good faith belief that use of the material in the manner complained of is not authorized by the copyright owner, its agent, or the law. Authority to Act: The information in this notification is accurate, and I state under penalty of perjury that I am authorized to act on behalf of the copyright owner.
\end{enumerate}

\section{Outcomes}

\begin{figure}[H]
    \centering
    \includegraphics[width=0.6\linewidth, trim={0 20 20 60}, clip]{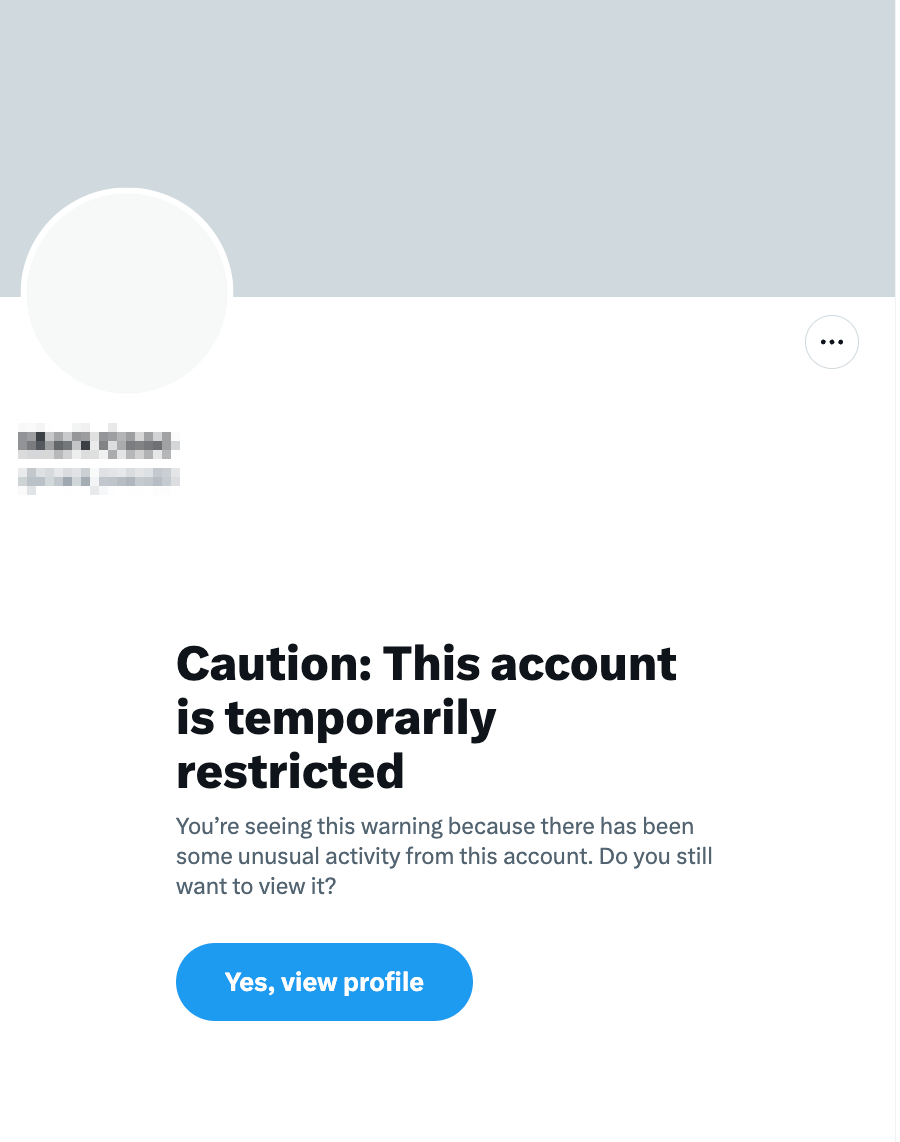}
    \caption{All five poster accounts reported under the DMCA condition were temporarily restricted. Poster accounts reported under the non-consensual nudity condition had no consequences.}
    \label{fig:sus-account}
\end{figure}

\begin{figure}[H]
    \centering
    \includegraphics[width=0.8\linewidth, trim={0 0 0 0}, clip]{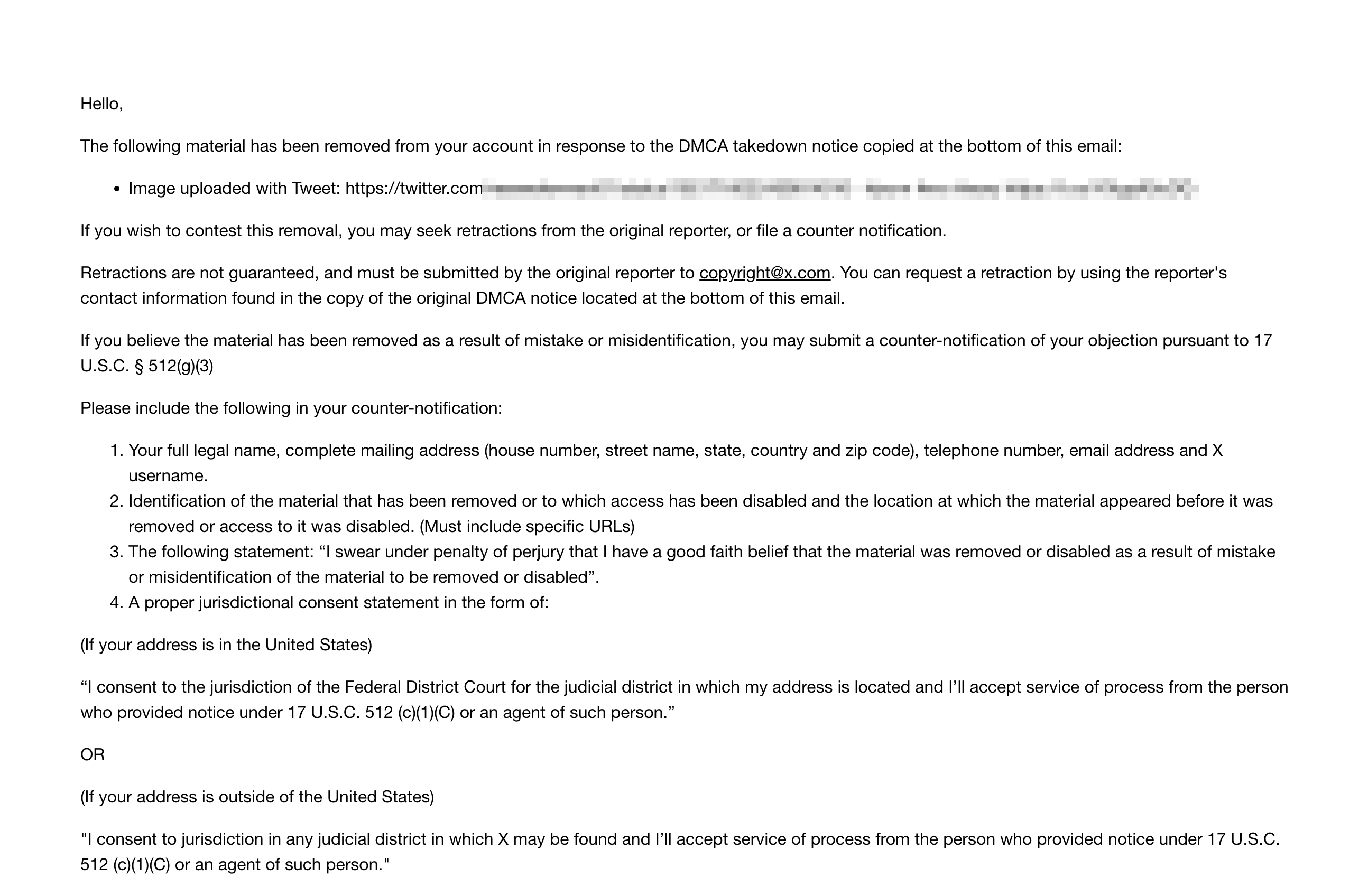}
    \caption{Email received with full DMCA reporter information. All five poster accounts reported with DMCA received this email from X.}
    \label{fig:sus-account-email}
\end{figure}

\section{Study materials}\label{study-materials}

\pixels

\neutralaccounts

\section{X content policies}\label{content-policy}

\begin{figure*}[h]
    \centering
    \begin{minipage}{0.43\linewidth}
        \centering
        \includegraphics[width=\linewidth]{figures/forms/dmca-form-1.pdf}
        \label{fig:cr-form-1}
    \end{minipage}
    \hfill
    \begin{minipage}{0.393\linewidth}
        \centering
        \includegraphics[width=\linewidth, trim={0 0 0 80}, clip]{figures/forms/dmca-form-2.pdf}
        \label{fig:cr-form-2}
    \end{minipage}
    \caption{Full copyright infringement report form on X. Copyright infringement is under intellectual property issues.\url{https://help.x.com/en/forms/ipi/dmca}}
    \label{fig:cr-form}
\end{figure*}

\begin{figure*}[h]
    \centering
    \includegraphics[width=0.53\linewidth,trim={0 0 0 0},clip]{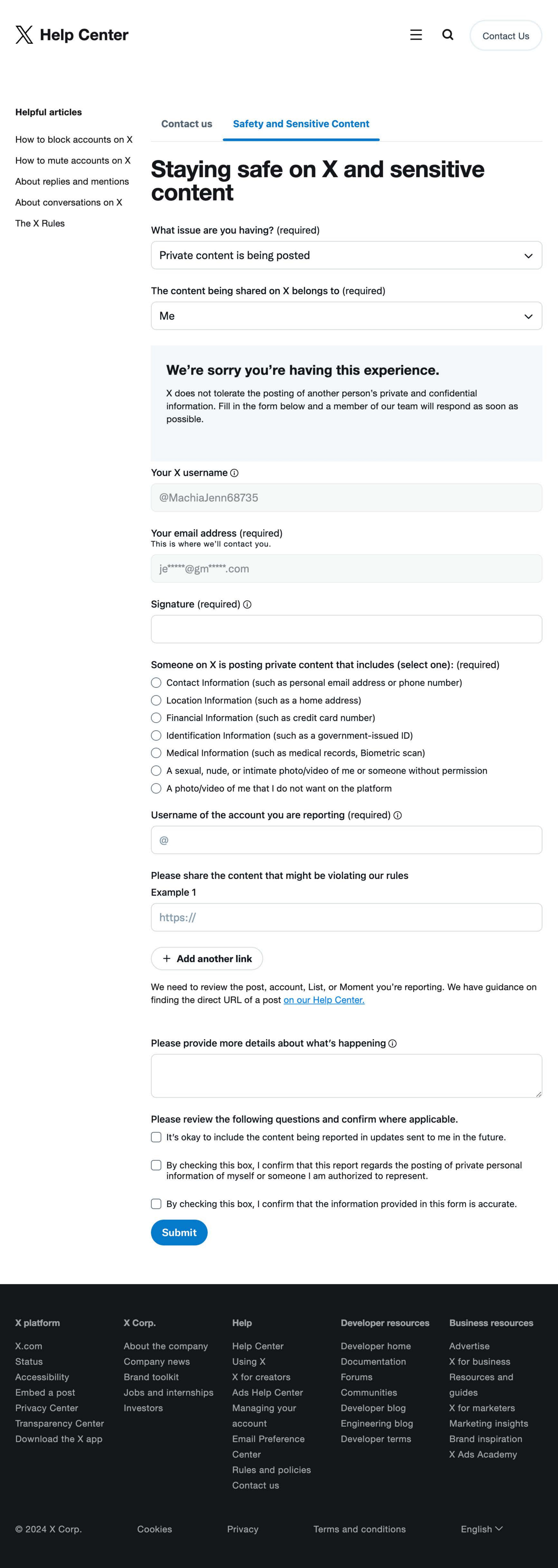}
    \caption{Full non-consensual nudity reporting form on X. Non-consensual nudity is under private content, which is part of safety and sensitive content. \url{https://help.x.com/en/forms/safety-and-sensitive-content/private-information}}
    \label{fig:safety-form}
\end{figure*}

\begin{table*}[h]
\centering
\fontfamily{cmss}\selectfont 
\begin{tabular}{p{6cm} p{3.5cm} p{4cm}}

\multicolumn{3}{c}{\textbf{NCIM:} non-consensual intimate media, the general issue we want to address} \\

\toprule
\textbf{Reporting condition} & \textbf{X's policy} & \textbf{Housed under} \\
\toprule
\textbf{NCN:} non-consensual nudity & Private content & Safety and sensitive content\\

\textbf{DMCA:} Digital Millennium Copyright Act & Copyright infringement & Intellectual property issues\\
\bottomrule
\end{tabular}
\caption{Glossary of frequently used terms, conditions, and how X handles these policies.}
\label{tab:terms}
\end{table*}

\ncimshortreport

\end{appendices}

\end{document}